\documentclass[aps,pre,twocolumn,groupedaddress,floatfix]{revtex4}
\usepackage{graphicx} 
\usepackage{graphics}
\usepackage{epsfig}
\usepackage{amsmath}
\usepackage[svgnames]{xcolor}
\usepackage{todonotes}
\usepackage{multirow}

\usepackage{outlines} 

\begin{document}
\title{Properties of packings and dispersions of superellipse sector particles}

\author{John Colt}

\altaffiliation [Rochester Institute of Technology, School of Physics and Astronomy]{}

\author{Lucas Nelson}

\altaffiliation [Haverford College, Department of Physics and Astronomy]{}

\author{Sykes Cargile}

\altaffiliation [Haverford College, Department of Physics and Astronomy]{}

\author{Ted Brzinski}

\altaffiliation [Haverford College, Department of Physics and Astronomy]{}

\author{Scott V. Franklin}

\altaffiliation [Rochester Institute of Technology, School of Physics and Astronomy]{}


\begin{abstract} \noindent 
Superellipse sector particles (SeSPs) are segments of superelliptical curves that form a tunable set of hard-particle shapes for granular and colloidal systems. SeSPs allow for continuous parameterization of corner sharpness, aspect ratio, and particle curvature; rods, circles, rectangles, and staples are examples of shapes SeSPs can model. We compare three computational processes: pair-wise Monte Carlo simulations that look only at particle-particle geometric constraints, Monte Carlo simulations that look at how these geometric constraints play out over extended dispersions of many particles, and Molecular Dynamics simulations that allow particles to interact to form random loose and close packings. We investigate the dependence of critical random loose and close packing fractions on particle parameters, finding that both values tend to increase with opening aperture (as expected) and, in general, decrease with increasing corner sharpness. The identified packing fractions are compared with the mean-field prediction of the Random Contact Model. We find deviations from the model's prediction due to correlations between particle orientations. The complex interaction of spatial proximity and orientational alignment is explored using a generalized Spatio-Orientational Distribution Area (SODA) plot. Higher density packings are achieved through particles assuming a small number of preferred configurations which depend sensitively on particle shape and system preparation. 

\end{abstract}

\maketitle

\section{Introduction}
A striking finding of studies of packings of elongated rods~\cite{Stokely,Desmond} or convex particles~\cite{Gravish2012,franklin2014extensional} is their entanglement, which can  produce bulk cohesion despite the absence of attractive inter-particle potentials.
A pile's particle shape-related resistance to tensile forces is called {\it geometric cohesion}, and the study of these systems represents a relatively new area of research in the broader field of granular materials.
Until recently there has been no general framework by which to systematically categorize particle shape or move smoothly from one shape to another to explore how important behaviors arise.
Thus, work exploring the role of particle shape has amounted to empirical phenomenology --- conclusions on staple-shaped particles, for example, are not easily related to that of rounded cubes. 

Critical granular volume fractions (e.g. random close and random loose packings) depend strongly on the symmetries of the constituent
particles~\cite{Donev990}. Particle anisometry can result in an
increased angle of repose~\cite{BEAKAWIALHASHEMI2018397,
  PhysRevE.82.011308}, even exceeding 90$^\circ$.~\cite{PhysRevE.82.011308,doi:10.1002/9781119220510.ch17}. Packings of staples, rigid and flexible rods, and star- and z-shaped particles can exhibit tensile strength~\cite{PhysRevE.101.062903}, with
relevance for aleatory design~\cite{doi:10.1063/1.5132809} and
strengthening granular materials under
strain~\cite{JaegerShear,Bares2019,PhysRevLett.120.088001}. Banana-shaped or bent-core rods are of broad interest to the liquid crystal community because of their rich 
phase-space~\cite{Yangeaas8829,PhysRevX.4.011024,FR950,RevModPhys.90.045004}. Semi-circular particles are a compelling, quasi-2D model system for the study of homodimerization and chirality-driven phase
separation~\cite{PhysRevE.94.022124,doi:10.1021/jacs.5b10549}, and
entropy approaches have been applied successfully to colloidal
crystals of a variety of shapes~\cite{Gengeaaw0514,
  Glotzer2007}. Recently, Ref.~\cite{PhysRevE.102.042903} used Monte Carlo
techniques to investigate packings of hard, circular arcs,
analytically identifying densest configurations which can then be
compared with simulations to identify the likelihood of their
appearing in bulk packings. 

Ref.~\cite{kornick2021excluded} introduced a new construct, the Super-ellipsoidal Sector Particle (SeSP), that can be parameterized to model a wide variety  of particle shapes, encompassing everything from stars to circles to disco-rectangles to staples by tuning variables in a parameterization of the Lam\'{e} curve $\left|x \right|^m + \left| y/A \right|^n = 1$:
\begin{equation}
\begin{bmatrix}
    x(\theta) \\
    y(\theta) \\
         \end{bmatrix} = 
\begin{bmatrix}
    \left|\cos\theta\right|^{2/n} \; A \; \textrm{sign}(\cos\theta) \\
    \left|\sin\theta\right|^{2/m} \; \textrm{sign}(\sin\theta) \\
        \end{bmatrix}.
\label{Eqn:para}
\end{equation}
$A$ is the particle aspect ratio, $m$ and $n$ the
superellipse degrees which control particle curvature, and the particle is restricted to a segment within $[\theta_{\rm min},\theta_{\rm max}]$. Ref.~\cite{kornick2021excluded} calculated the excluded area for a characteristic set of SeSPs and mapped the complicated relationship between orientation and relative position for non-overlapping configurations of a pair of SeSPs. In this work, we study the relationship between these features of the configurations of isolated pairs of SeSPs and the multiscale structural features of dense dispersions and packings.

\section{Methodology}

\begin{figure*}[tp!]
    \centering
        \includegraphics[width=0.20\linewidth]{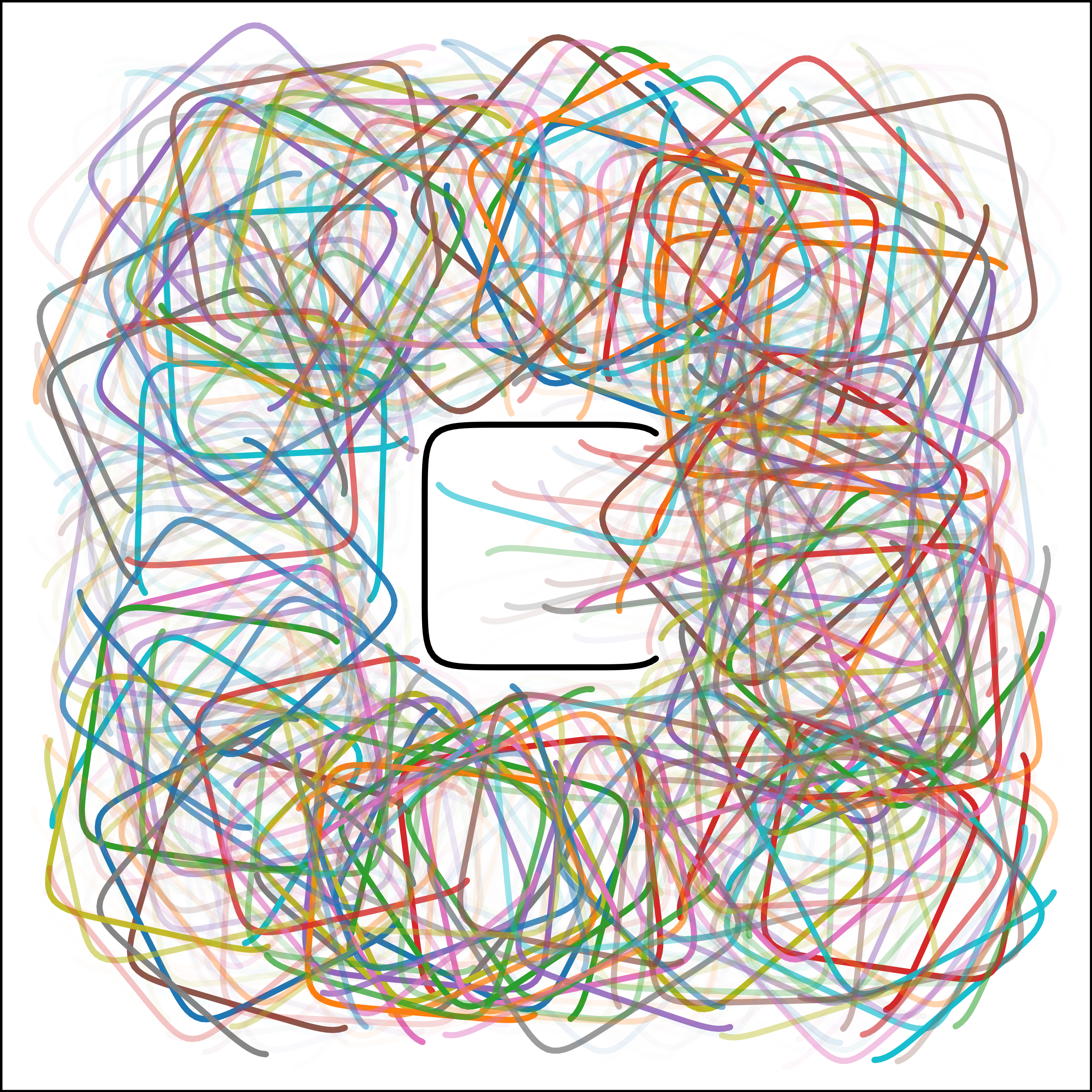}\hskip 0.1in
        \includegraphics[width=0.20\linewidth]{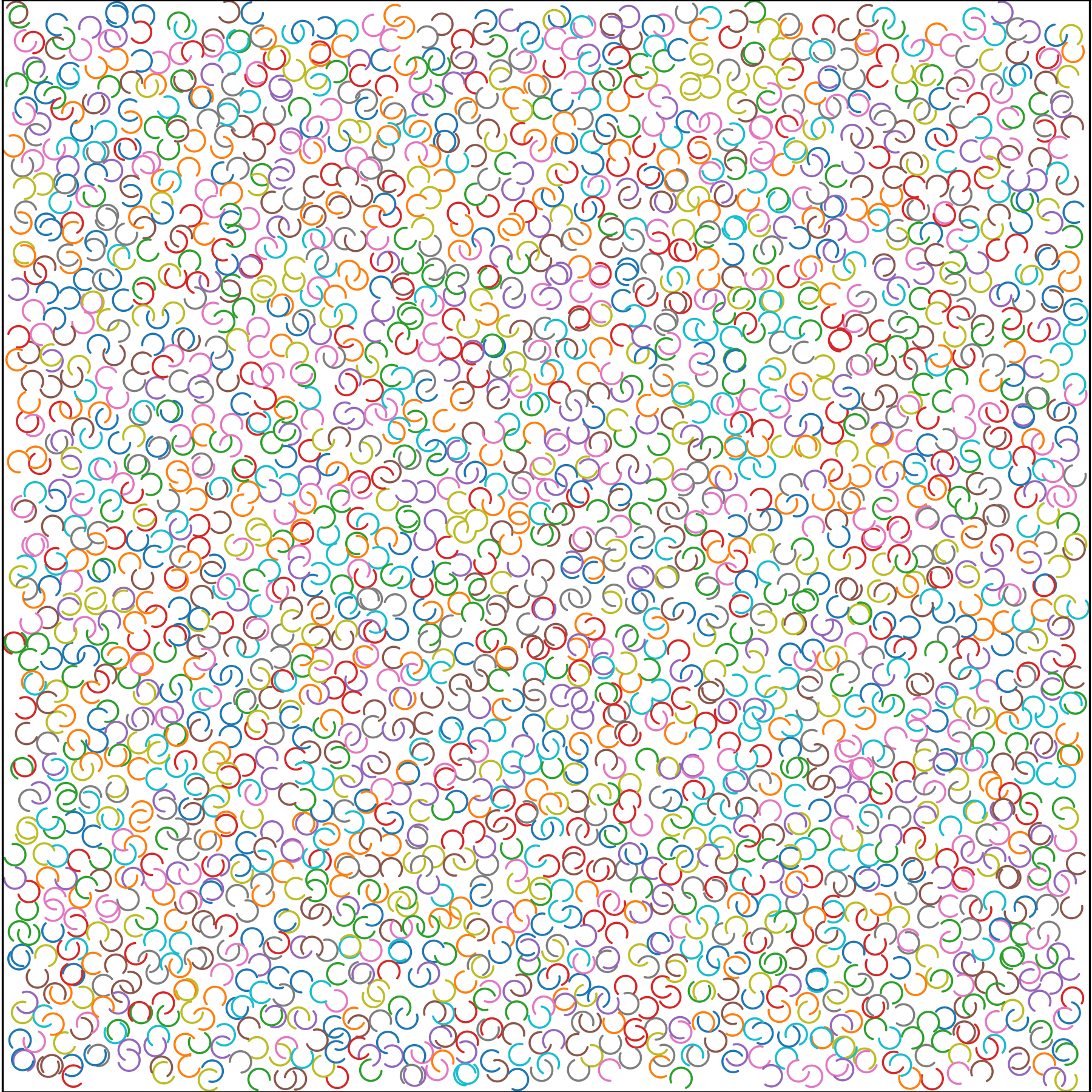}\hskip 0.1in
        \includegraphics[width=0.20\linewidth]{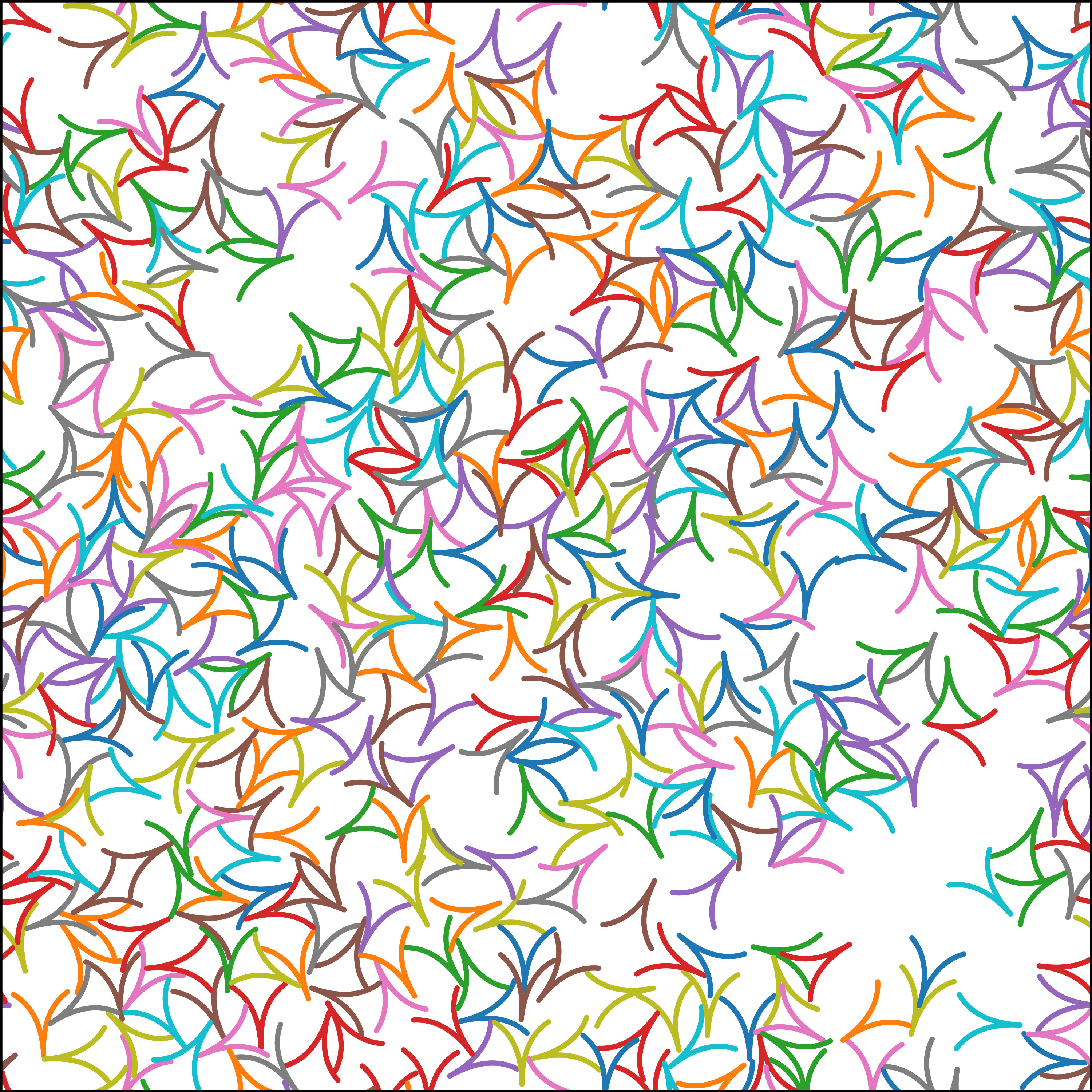}\hskip 0.1in
        \includegraphics[width=0.20\linewidth]{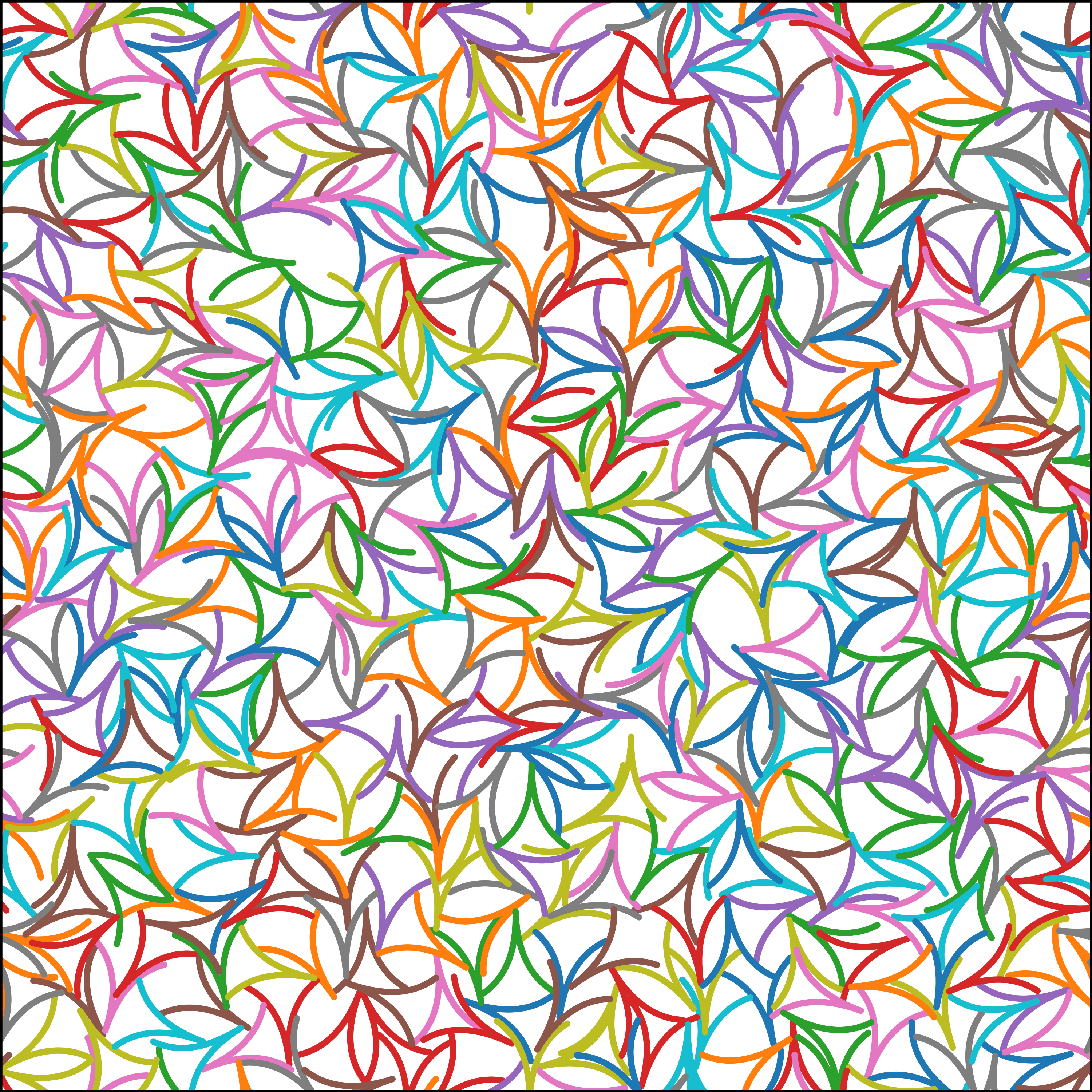}
    \caption{{\it Left}: Sample particle placements from a pairwise Monte Carlo simulation.
    {\it Center Left}: A dispersion of $\sim 3000$ SeSPs created through a Monte Carlo simulation.
    {\it Center Right}: A loose packing of $500$ SeSPs created through a Molecular Dynamics simulation.
    {\it Right}: A close packing of $500$ SeSPs created through a Molecular Dynamics simulation.}
    Different SeSP parameters chosen to represent the diversity of particle shapes.
    \label{fig:prdisppckpck}
\end{figure*}
The particle-pair metrics introduced in ref.~\cite{kornick2021excluded} are based on a straightforward Monte Carlo method (described in more depth in ref.~\cite{PhysRevE.67.051302}) which proceeds as follows:
(1) a particle is fixed at the origin and (2) a second particle is placed at a random location and orientation and checked for overlap with the first. If there is no overlap, the configuration is recorded. The second particle is then moved to a new (random) location and orientation and the process repeats.

Dispersions are created with a generalization of the same Monte
Carlo (MC) method: Particles are placed one at a time at random
locations and orientations in an extended space and checked for overlap with previously placed particles. If an overlap is present, the newly placed particle is removed. In contrast to the pairwise study, the randomly placed particle is not removed before the next iteration. Instead, the process then repeats for a newly-located and oriented particle and continues until the desired packing fraction has been reached or no new particles can be placed without overlap (within some long computational time-frame).

In order to extend our study to rigid packings, we also utilize molecular dynamics (MD) simulations, using the well-established LAMMPS MD software~\cite{LAMMPS}. In the simulations, a given number of particles are placed randomly and at low density within a 2d space and slowly expanded.
Particles are checked for interaction with neighbors at each expansion step, with interactions following an Hertzian repulsion, and evolve according to Newton's Laws.
The simulations are over-damped in order to achieve rest and suppress numerical noise; expansion is slow enough to ensure quasi-static behavior.
The transition to Random Loose Packing (RLP), the concentration at which the system transitions from a liquid dispersion to a rigid packing, is marked by the onset of jamming, which we identify by the rigidity percolation criterion.

Rigidity percolation defines the state of the system in terms of particle clusters which are rigid independent of the rest of the packing.
We assess the rigidity of the internal structure with the Pebble Game algorithm~\cite{JACOBS1997346}, modifying code published by Silke Henkes~\cite{PhysRevLett.126.088002}.
We identify jamming at the concentration at which the largest rigid cluster spans the system.
An advantage of this analytic approach is that the scaling of these rigid structures as the system approaches jamming is a well-established metric by which to characterize the state of the system~\cite{PhysRevE.53.3682,PhysRevLett.114.135501,PhysRevLett.116.028301,PhysRevLett.121.188002,PhysRevLett.123.058001,PhysRevX.9.021006,PhysRevE.99.012123}, allowing us to connect to a broad library of prior research on particulate matter. We can also continue the quasistatic expansion of the particles until particle interactions become aphysical (e.g. particles are forced to intersect). We label highest density state of a packing without such aphysical particle interactions as Random Close Packing (RCP).
 
Examples of dispersions and packings prepared by each of the procedures described above are shown in Fig.~\ref{fig:prdisppckpck}.

In this work, we investigate SeSPs with aspect ratio $A=1$ and equal superellipse degrees $m=n$, and choose segments $[\theta_{min}, \theta_{max}]$ such that the opening in the SeSP subtends particular polar angles $\Psi$ measured relative to the center of curvature and centered on the x-axis.

\section{Results}
\subsection{Random Loose and Close Packings}
\begin{figure}[ht]
\includegraphics[width=1\linewidth]{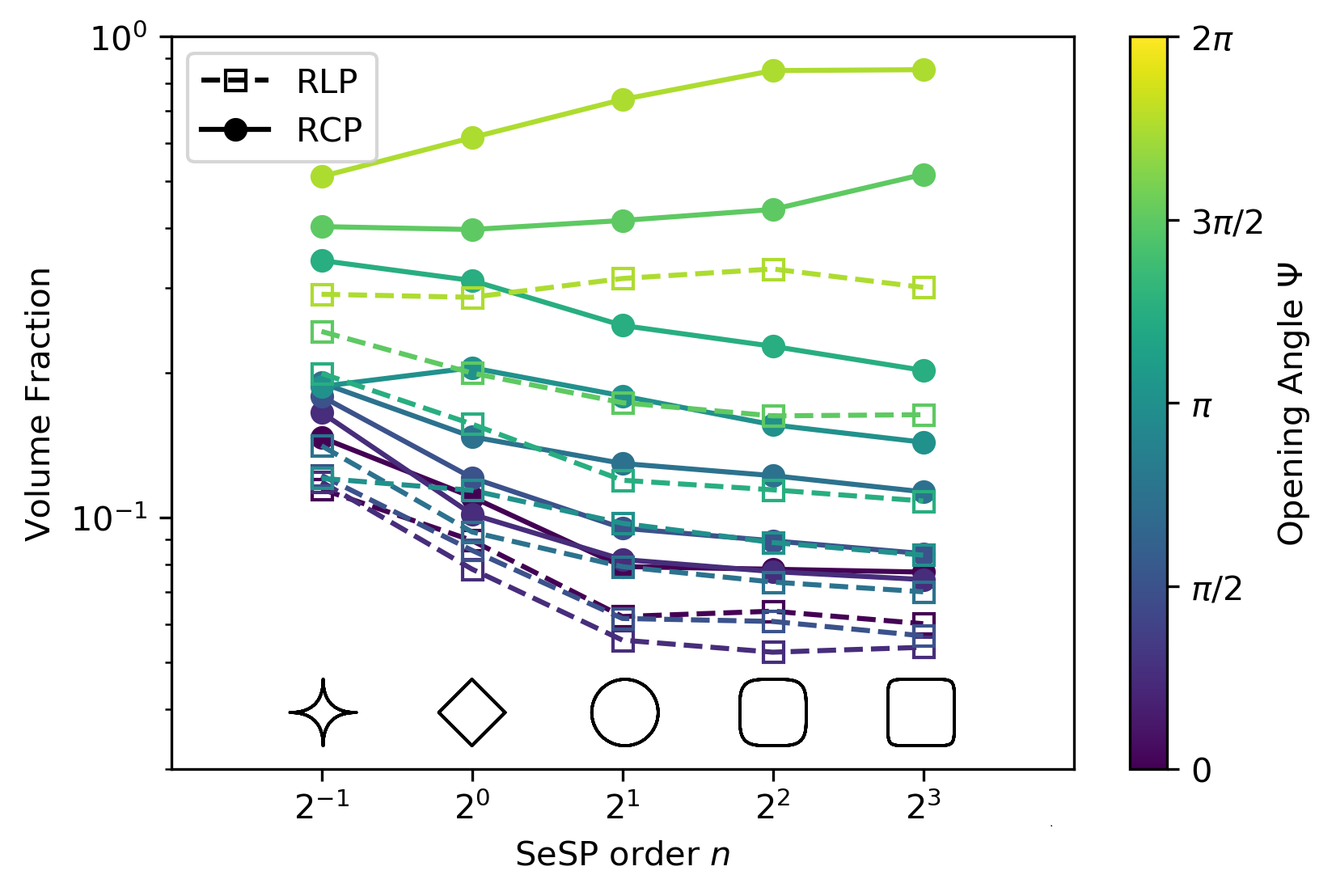}
\caption{\label{packings}  Critical packing fractions plotted vs SeSP exponent $n$, showing the dependence on corner sharpness. Colors indicate the opening angle of the SeSPs, line and marker types indicate whether the volume fraction is measured at Random Loose or Random Close Packing, and inset at bottom shows representative SeSPs of the various orders included with zero opening angle.
}
\end{figure}

The values of the Random Loose and Random Close Packing fractions for the chosen subset of SeSPs as functions of exponent for different opening apertures $\Psi$ are shown in Fig.~\ref{packings}. The opening aperture is indicated by color, ranging from purple ($\Psi=0$, a closed particle) to light green ($\Psi\rightarrow 2\pi$, an almost completely open particle). As expected, the critical volume fractions increase with opening aperture, as particles with larger openings allow entanglement configurations that result in denser packings. With the exception of only the most open particle shapes, the critical packings, both Random Loose (open boxes) and Random Close (closed circles) general decrease with corner sharpness (SeSP order $n$).
This is due to a combination of factors, including the additional rotational constraints which sharp corners (and correspondingly straight sides) impose, as well as the fact that higher-order superellipses with the same thickness occupy a smaller fraction of their enclosed area.

\begin{figure}
    \centering
    \includegraphics[width=1\linewidth]{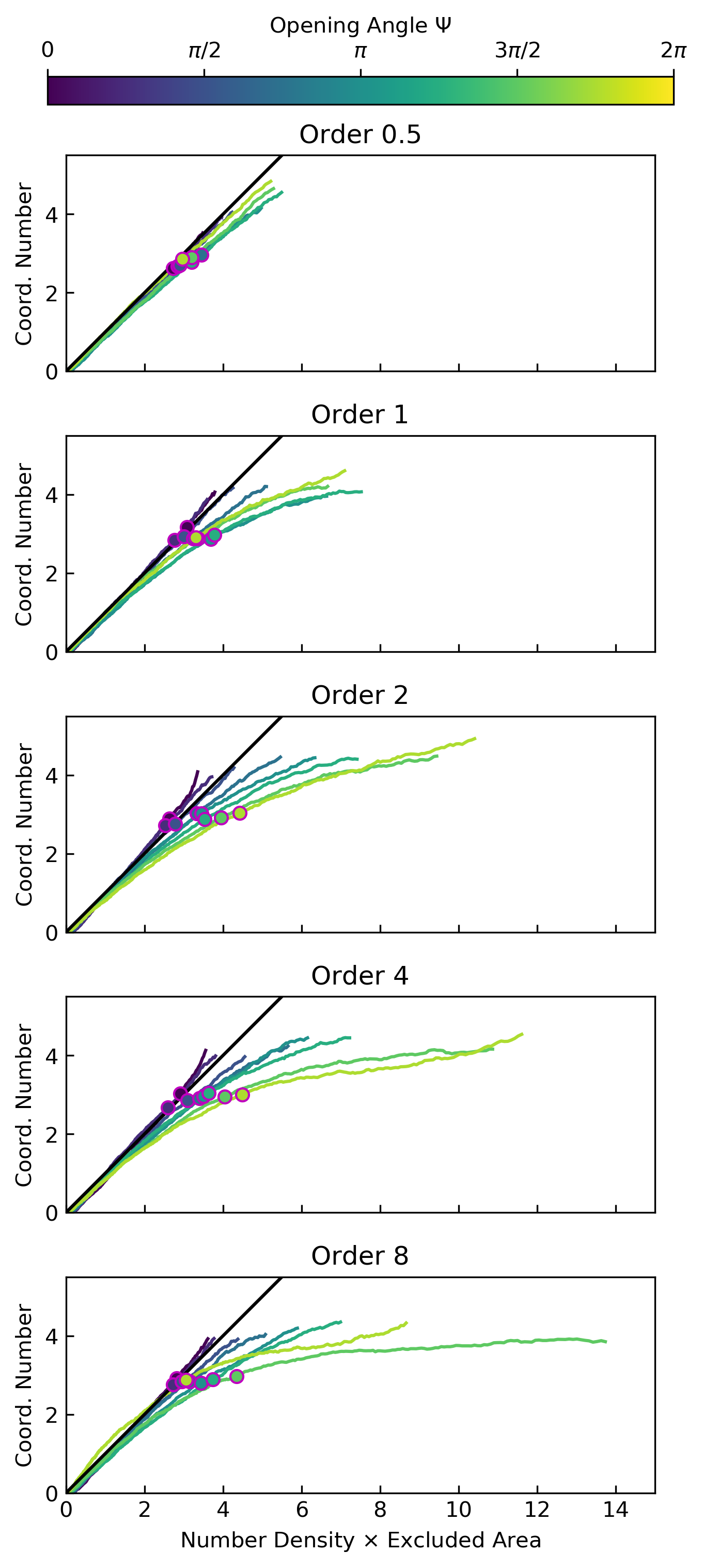}
    \caption{Mean coordination number plotted against the product of excluded area and number density. Each colored line represents a full Molecular Dynamics simulation with a different shape of SeSP, with hollow circles at random loose packing and filled circles at random close packing. Black diagonal lines represent the prediction of the Random Contact Model, which gives that the two quantities should be equal for an uncorrelated random packing.}
    \label{fig:RCMPhi}
\end{figure}

The Random Contact model \cite{Philipse1996} posits that critical packing fractions (or associated number density) scale inversely with the excluded area, and that the product of the number density and excluded area should equal the mean coordination number of the packing.
This mean-field model rests on the assumption that there is no positional or orientational correlation between nearby particles, so any significant deviation from this prediction implies a substantially ordered packing.
We calculate the particle excluded area for each shape from pairwise simulations as in \cite{kornick2021excluded}, and identify the mean coordination number at every step of each molecular dynamics simulation by counting contacts.
We then plot the mean coordination number vs. the product of the excluded area and the number density of every step in the simulations to test the model's predictions. This is shown in Fig.~\ref{fig:RCMPhi}, with black lines drawn to indicate the prediction that the two quantities should be equal.
This figure indicates that the model is very accurate for low packing fractions, where contacts between particles are relatively sparse, but begins to diverge from reality for many shapes of SeSP close to RLP (indicated by hollow circles on the plot) and diverges much more significantly as the simulations approach RCP.
We also find that the accuracy varies significantly based on particle shape; higher-opening-angle SeSPs tend to have fewer contacts at a given packing fraction than the model would predict, with a larger deviation for larger opening angles. However, this tendency does not appear to hold for order-1/2 SeSPs, which remain highly angular even at large opening angles.

This result is a plausible conclusion from the packing preparation, and suggests that Random Loose packings form with less rearrangement driven by particle-particle interactions which would encourage particle orientational correlations. Random Close Packing, however, occurs well after particles have first come into contact with neighbors, and the packing process involves particles rearranging around one another to explore more dense configurations. These rearrangements result in correlations between neighbors and therefore violate the Random Contact Model's assumptions. The same effect was seen in quasi-two-dimensional, experimental packings of rods~\cite{Stokely}, a particle geometry equivalent to SeSPs of $n>>2$ and $\Psi>3\pi/2$.
These results highlight the need to understand particle-scale mechanics in order to understand the material's bulk structure and response. SeSPs provide a framework by which to study how these mechanics influence local structure as a function of particle geometry. In particular, as our system deviates further from the random contact model, we may expect growing correlations between particle pair orientation and position.

\subsection{Spatial and Spatio-orientational Correlations}

\begin{figure*}
    \centering
        \begin{minipage}{0.4\linewidth}
            \begin{tabular}{c}
                \includegraphics[width=0.95\linewidth]{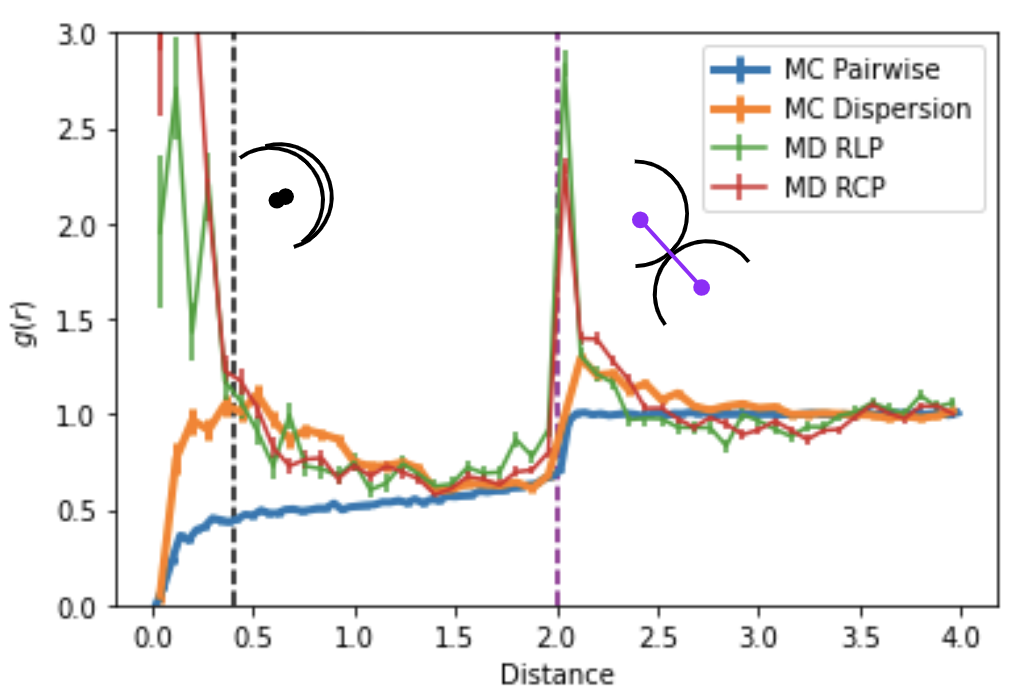}
                \\
                \includegraphics[width=0.95\linewidth]{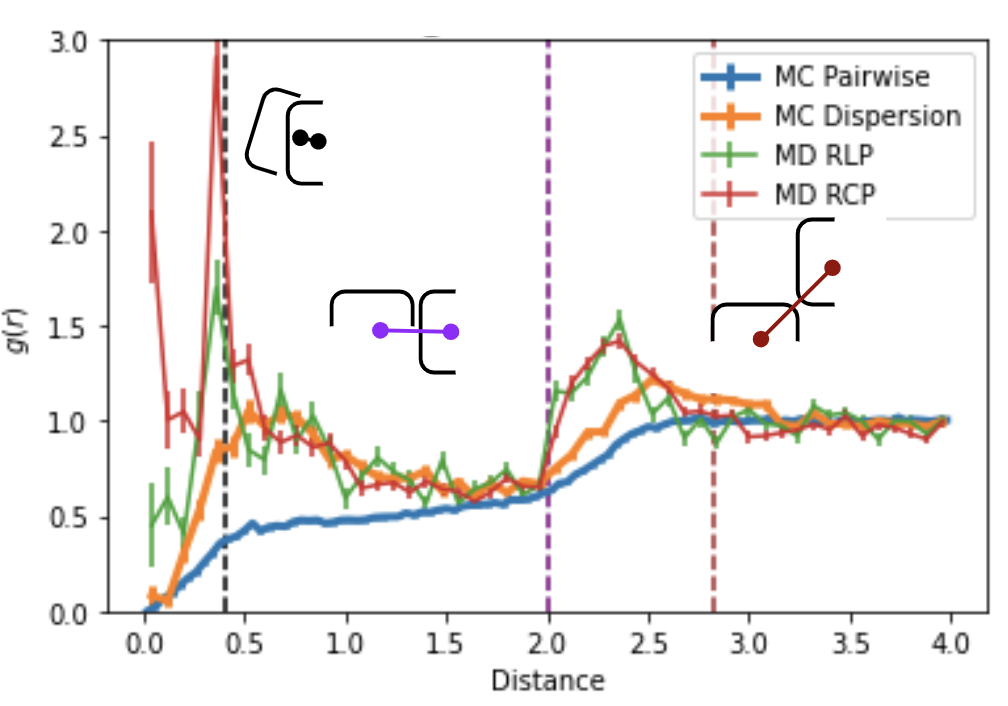}
            \end{tabular}
        \end{minipage}
        \begin{minipage}{0.3\linewidth}
            \begin{tabular}{c}
                      \\
                \includegraphics[width=0.8\linewidth]{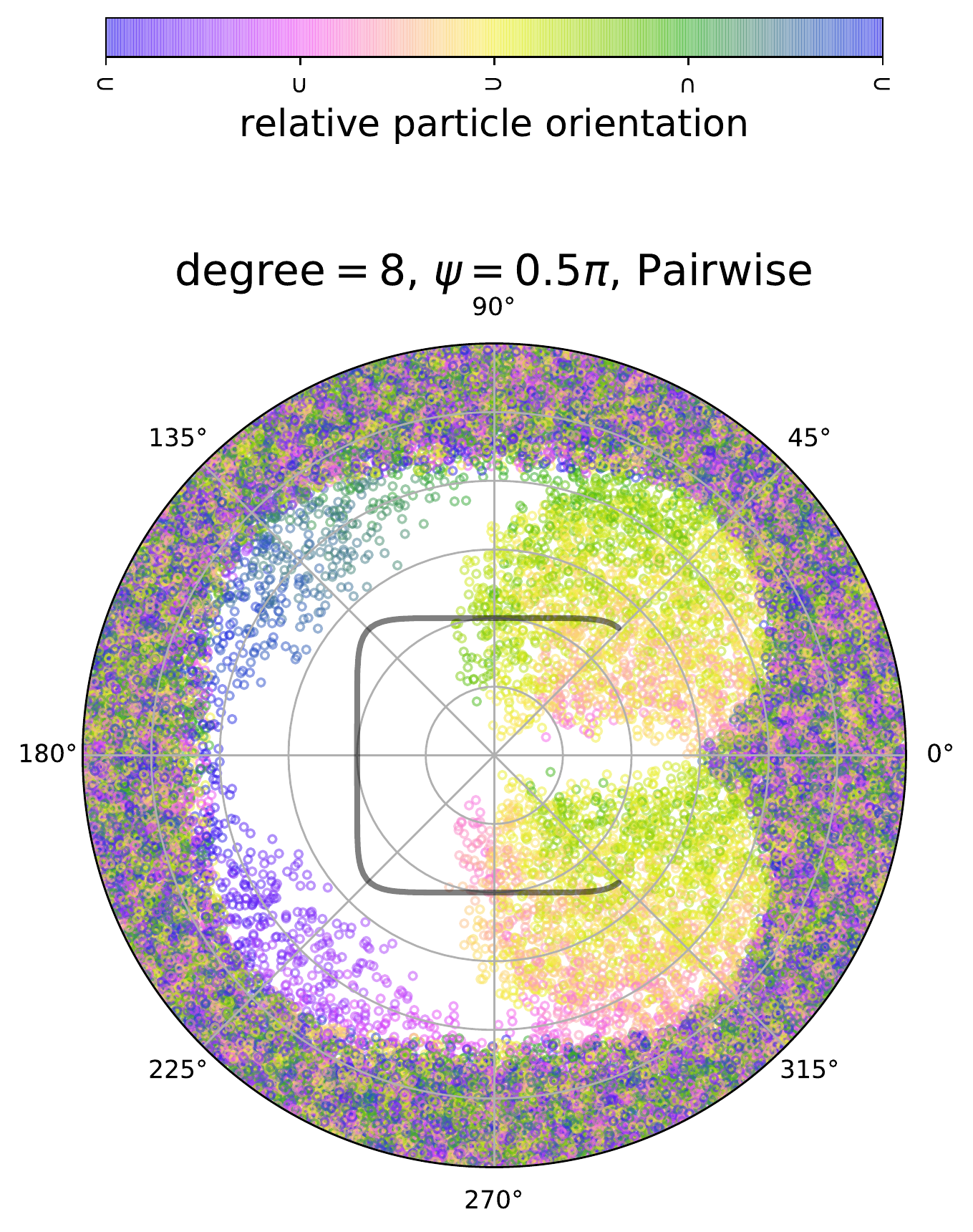}
            \end{tabular}
        \end{minipage}
        \begin{minipage}{0.2\linewidth}
            \begin{tabular}{c}
                {\large Orientation}
                     \\
                \includegraphics[width=\linewidth,trim={0 0 0 25},clip]{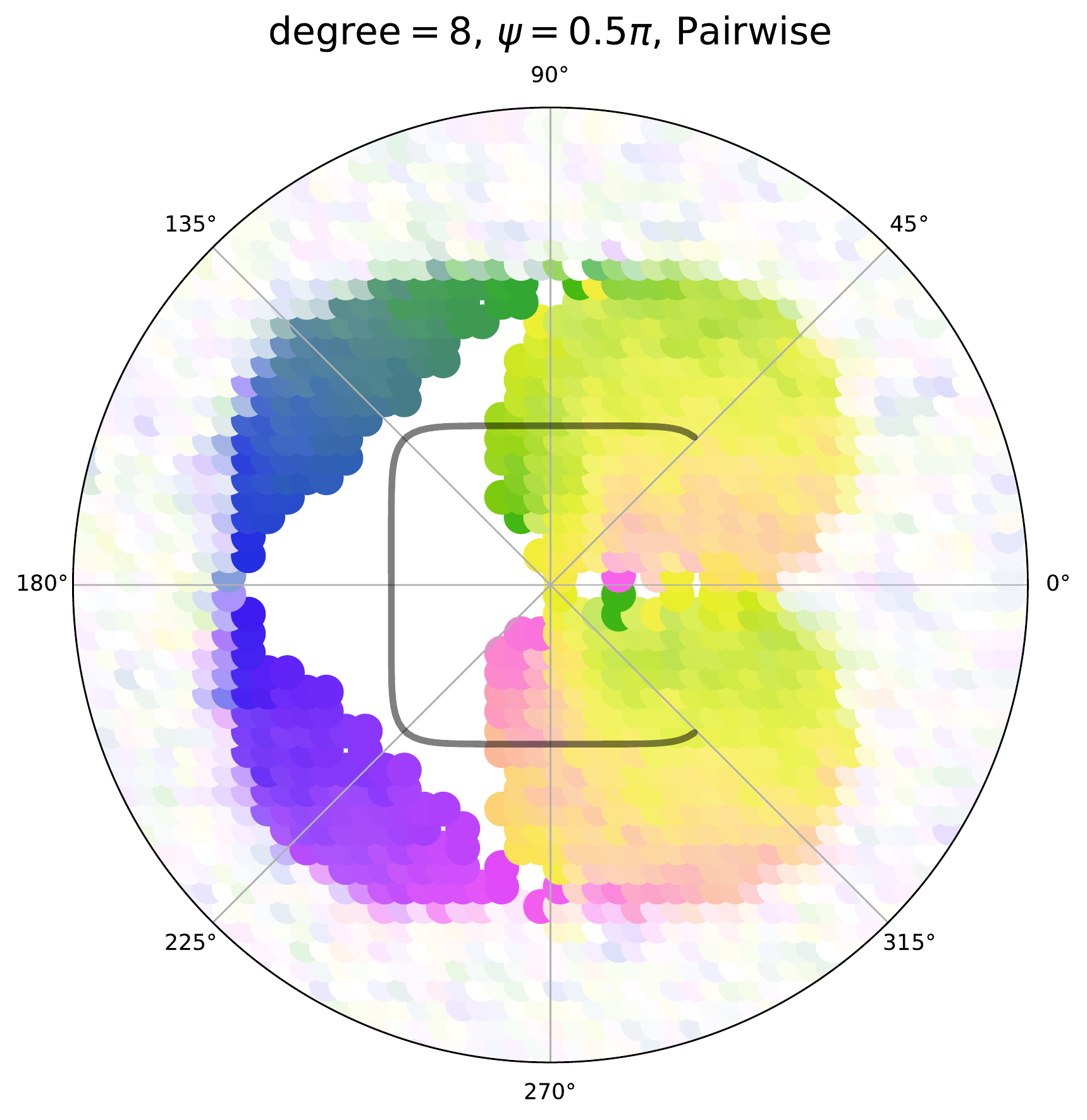} 
                     \vspace{1em}
                     \\
                {\large Density}
                     \\
                \includegraphics[width=\linewidth,trim={0 0 0 25},clip]{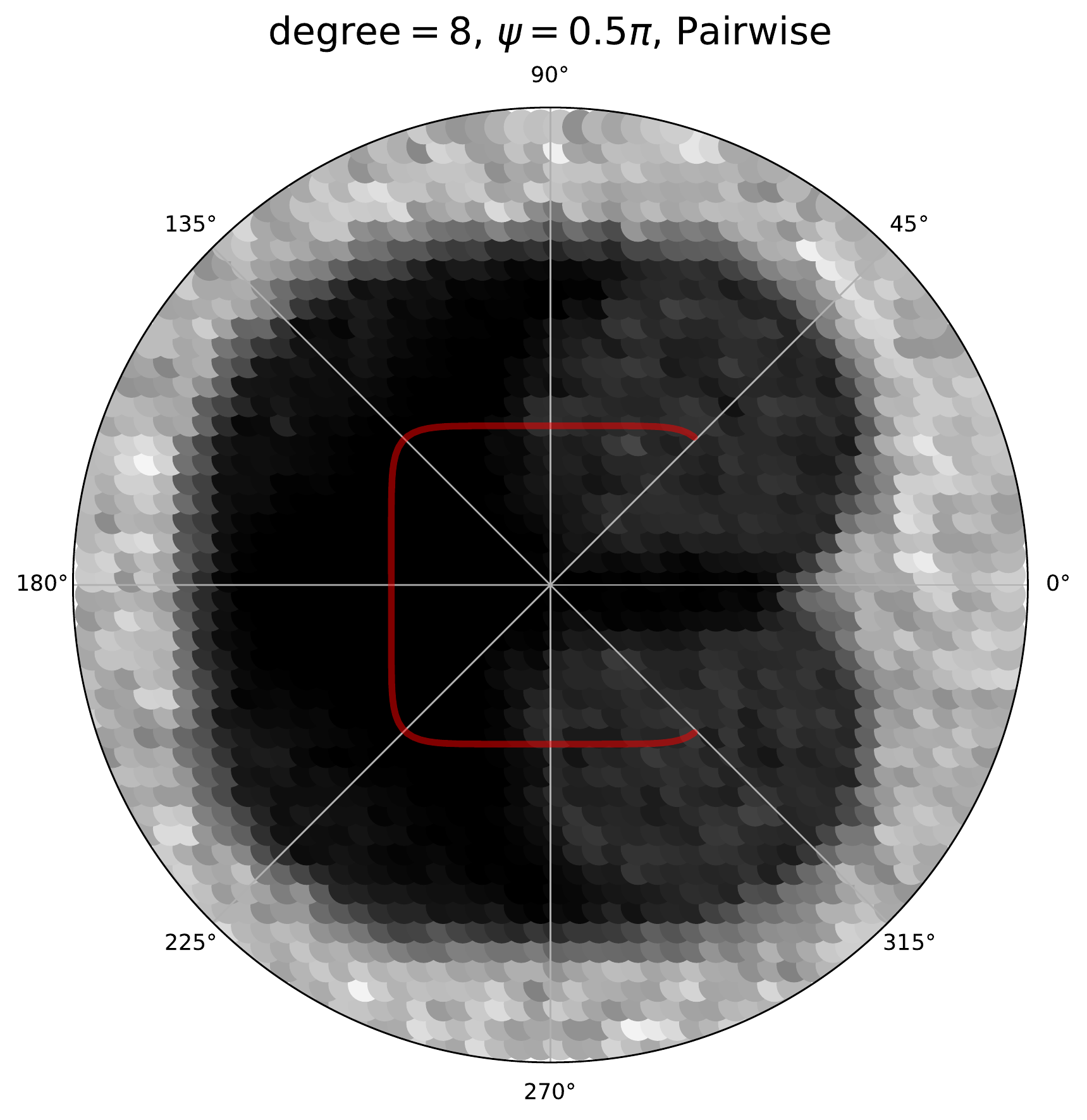} 
                     \\
                \includegraphics[width=\linewidth,trim={0 25 0 0},clip]{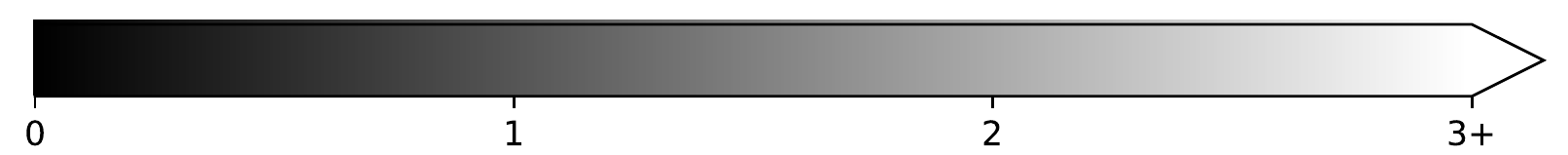}\\
                {$n/n_\textrm{mean}$}
            \end{tabular}
        \end{minipage}
    \caption{{\it Left}: Pair correlation function $g(r)$ for semi-circular (top) and staple-shaped (bottom) SeSPs. The distributions show a step at (semi-circles) or slightly beyond (U-shaped) $r=2$, the distance at which particles can always be placed regardless of orientation. Peaks around these values are  seen in both dispersions and packings, indicating preferential locations for more densely packed particles and identifying a characteristic length scale for the collections. {\it Middle/right}: Three methods of visualizing the positions and orientations of particles relative to a central sample particle. \emph{Middle:} a SODA plot, in which each point represents the center of a nearby SeSP and the color indicates relative orientation. \emph{Upper Right:} an orientation map, in which pixel hue corresponds to the average orientation of nearby particles within a certain distance of that pixel, and pixel saturation indicates the standard deviation of those orientations. Full saturation corresponds to a standard deviation of zero, and zero saturation corresponds to a standard deviation equal to or greater than the standard deviation of a uniform distribution. \emph{Lower Right:} a density map, in which pixel lightness corresponds to the number of nearby particle centers within a certain distance of that pixel, normalized by the mean number of nearby particles.}
    \label{fig:gofrDeSODA}
\end{figure*}

We compute the  pair-correlation function
\begin{equation}
    g(r) =\frac{A}{N^2}\frac{\sum_{i}^{N}n_i(r,r+dr)}{2\pi rdr,}
    \label{eq:gr}
\end{equation}
which measures the probability of finding another particle's center of curvature as a function of the distance $r$ from another particle's center of curvature. The numeric calculation of $g(r)$ is an ensemble average over all pairwise combinations of particles, and is normalized to $g(r\rightarrow\infty)=1$. Fig.~\ref{fig:gofrDeSODA}(left) shows the resulting distribution for 
for MC pairwise (blue) and dispersions (orange) and MD Random Loose (green) and Close (red) packings of SeSPs with $\Psi=\pi$ and $n=2$ (top) or $n=8$ (bottom) particles.

For SeSPs with opening angle $\Psi=\pi$, the particle shape and finite thickness prevents particles' centers-of-masses from overlapping and so $g(r\rightarrow 0) = 0$. The pairwise data show a smooth, rapid rise from 0 before a slowly increasing plateau for $0.5<4<2$ (in units of particle radius of curvature). The plateau for pairwise particles arises because, for the particular shape chosen, the number of orientations that do not result in overlap does not appreciably increase as the particles move apart. This changes at $r=2$, where suddenly all orientations are allowed, and the distribution jumps again to $1$. for $r>2$ all orientations are allowed and $g(r>1)\equiv 1$. The pairwise distribution represents the least dense configuration and a lower bound for dispersions and packings.

This is seen in the the comparison with the resulting curve from dispersions. This curve rises from zero to a peak at $r\approx 0.5$, representing the increased probability of close nearest neighbors. The dashed line at $r=0.4$ corresponds to the spooning/nesting orientation shown in Fig.~\ref{fig:gofrDeSODA}(top left). As the dispersion is populated, these orientations are used to fit additional particles into the small spaces. This peak is still larger for MD packings at Random Loose and Close packings, and is shifted slightly leftward to smaller values of $r$ as a consequence of the rearrangements enforced by the packing procedure. During this process, particles will re-orient very slightly to explore configurations that allow larger number densities, increasing the probability of finding neighbors at the nearest possible separations. In the corresponding plot for SeSPs with $n=8$, Fig.~\ref{fig:gofrDeSODA}(bottom left), the peaks are shifted rightward toward larger $r$ (compared with semi-circular particles), as the larger square diagonal further separates particles. The peaks in the packings for distributions again appear at slightly smaller separation distances as the packing protocol allows for optimization of packing.

The complex interaction between particle location and orientation was represented in Ref.~\cite{kornick2021excluded} with {\it Spatio-Orientational Distribution Area} or ``SODA'' plots.
A characteristic SODA plot is shown in Fig.~\ref{fig:gofrDeSODA}(center-right).
In these plots, locations (dots) of particles relative to the (shown) reference particle are summed over every choice of reference particle in the system.
Dots are colored based on the relative orientation of the SeSPs, and the eye quickly identifies several distinct features.
The space behind the reference particle is primarily comprised of particles oriented in the same direction (purples and blues) as the reference particle, creating a ``spooning'' or ``nesting'' region.
Around each endpoint of the reference particle is a circular region of particles primarily in the opposing direction of the reference particle (light yellows, greens, and pinks), an {\it entanglement} region.
For $\Psi<\pi$ a small white area where these two circular regions overlap indicates an {\it excluded} region where particles cannot be placed regardless of orientation.
This excluded region also extends to a sizable area inside the nesting region where the opening is too small to permit nesting behavior.
Finally, at the boundaries of each SODA plot, particles can be placed regardless of orientation, seen in the presence of dots of all colors.

\begin{figure*}
    \begin{tabular}{cccc}
        isolated particle pairs & non-rigid dispersion & random loose packing & random close packing \\
        \includegraphics[width=0.24\linewidth,trim={0 0 0 27},clip]{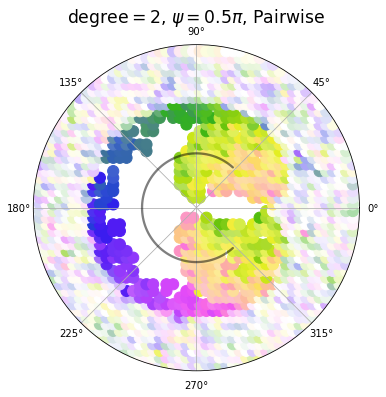} &
        \includegraphics[width=0.24\linewidth,trim={0 0 0 27},clip]{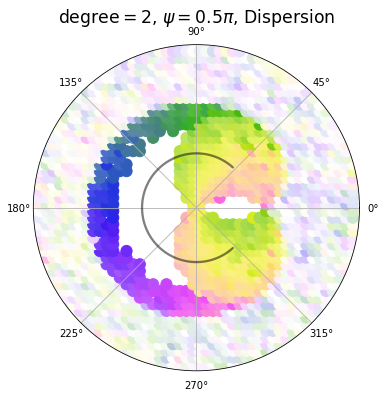} &
        \includegraphics[width=0.24\linewidth,trim={0 0 0 27},clip]{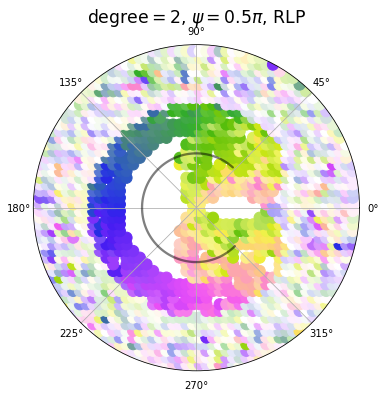} &
        \includegraphics[width=0.24\linewidth,trim={0 0 0 27},clip]{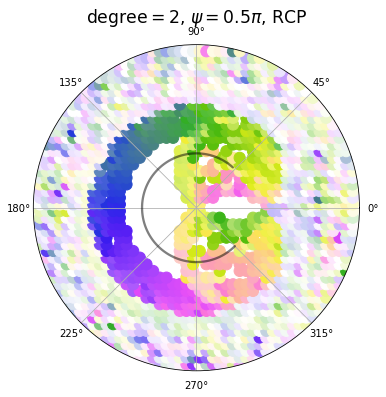} \\
        \includegraphics[width=0.24\linewidth,trim={0 0 0 27},clip]{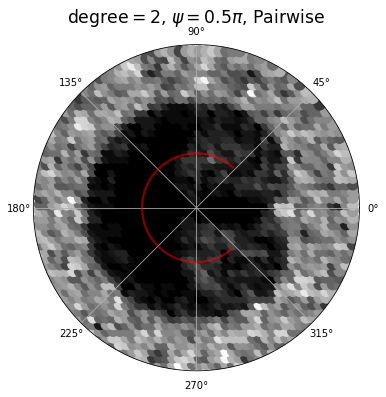} &
        \includegraphics[width=0.24\linewidth,trim={0 0 0 27},clip]{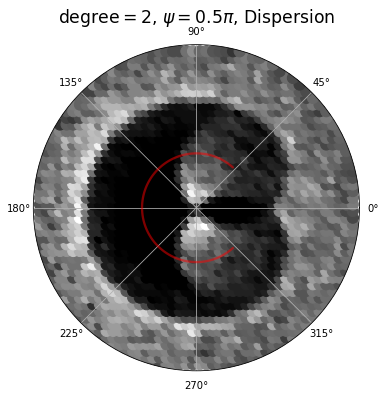} &
        \includegraphics[width=0.24\linewidth,trim={0 0 0 27},clip]{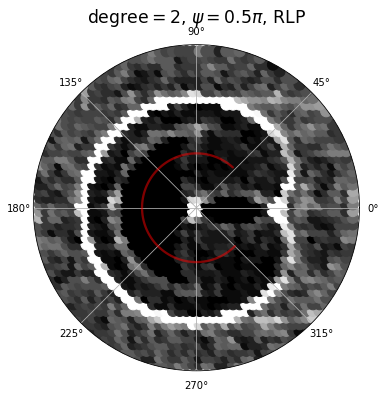} &
        \includegraphics[width=0.24\linewidth,trim={0 0 0 27},clip]{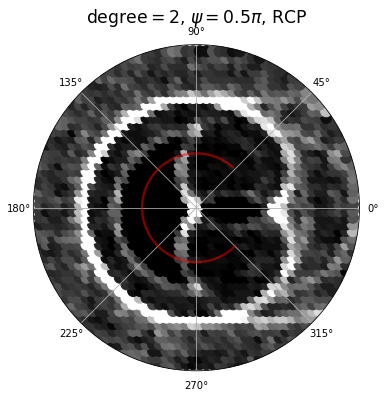}
    \end{tabular}
    \caption{SODA plots decomposed into orientation maps (top) and density maps (bottom) are shown for $n=2, \Psi=\frac{\pi}{2}$ SeSPs generated (left-to-right, as indicated) as isolated particle pairs, as a non-rigid dispersion, at random loose packing, and at random close packing.}
    \label{fig:deSODA-Compare}
\end{figure*}

The features described above capture the general traits of all SODA plots, whether generated for isolated particle pairs, non-rigid dispersions, or rigid packings, and for varied SeSP parameters.
The contrasts between these SODA plots offers a hint as to how local structure emerges as multi-particle interactions play an increasingly important role in constraining the availability of particular pairwise configurations.
In order to disentangle orientational correlations and density correlations that emerge in the SODA plots, we can decompose a SODA plot into a locally-averaged colormap of mean relative orientation, where the saturation of the colormap is inversely related to the standard deviation of the relative orientation, and a greyscale heatmap illustrating the probability of finding another SeSP centered at a particular point relative to the reference SeSP.
An example of the decomposition is shown in Fig.~\ref{fig:gofrDeSODA}(right), and decomposed SODA plots for $n=2$, $\Psi=\frac{\pi}{2}$ SeSPs for each of isolated particle pairs (left), non-rigid dispersions (center-left), and rigid packings (RLP center-right, RCP right) are shown in Fig.~\ref{fig:deSODA-Compare}.

\begin{figure}[ht]
\includegraphics[height=1.7in]{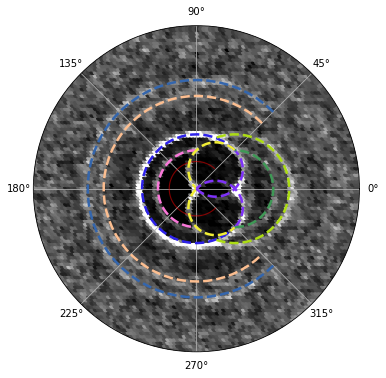}
\includegraphics[height=1.5in,clip]{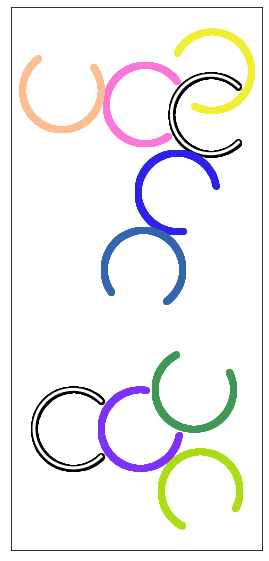}

\caption{\label{fig:SODAExplainer} (Left) Extended SODA density map for $n=2, \Psi= \tfrac{\pi}{2}$ SeSPs, with regions of high density colored. (Right) Schematic representation of the first- and second-order contacts which generate these high-density regions, shown relative to two black reference SeSPs, with colors corresponding to the high-density region they represent.}
\end{figure}

In the decomposed SODA plots, we observe that both density and orientational correlation tend to change abruptly at the boundaries between spooning, nesting, and non-interacting regions. In all of the systems analyzed, the orientational correlation drops off abruptly at the edges of the interacting regions and typically transitions sharply in direction between the nesting region and the entangled region, indicating the sudden change in permitted angles. The behavior of the density at these boundaries varies between the different methods of producing the packings in the same way as $g(r)$; the rigid packings feature sharp peaks in density along these boundaries due to particles rearranging to pack more tightly, while the non-rigid dispersions feature more gradual peaks and the isolated particle pairs simply drop off in density. The density maps for rigid packings also sometimes feature peaks in density outside of the interaction zones due to thee-or-more particle interactions; for instance, the density plot for random close packing in Fig.~\ref{fig:deSODA-Compare} features peaks to the right of the entanglement zones due to interactions with a SeSP in the opening of the reference SeSP, and a zoomed-out version of this density plot in Fig.~\ref{fig:SODAExplainer} illustrates more distant peaks which relate to other types of three-particle interactions.

\section*{Conclusion}
We have explored the statistics of multi-particle configurations of Super-ellipsoidal Sector Particles, a framework that can approximate an extraordinarily broad class of 2D particle shapes. We have measured the critical random (loose and close) packing fractions and their correlation to the calculated excluded area, finding discrepancies with the predictions of the mean-field Random Contact Model. These discrepancies are explained by the complex relationship between spatial proximity and orientational alignment, allowing particles to pack more densely through coordinated spatio-orientaitonal positioning with neighbors.

\bibliographystyle{unsrt}
\bibliography{proposal,references,NickBib,antbib,masterbib,refs,bib}

\end{document}